\documentclass[superscriptaddress,twocolumn,showpacs,pre,floatfix]{revtex4-1}

\usepackage{amsmath}
\usepackage{color}
\usepackage{tabularx}
\usepackage[dvips]{graphicx}

\begin{document}

\title{Ising Spin Glasses and Renormalization Group Theory : the Binder cumulant}

\author{P. H.~Lundow} \affiliation {Department of Mathematics and
  Mathematical Statistics, Ume{\aa} University, SE-901 87 Ume{\aa}, Sweden}

\author{I. A.~Campbell} \affiliation{Laboratoire Charles Coulomb
  (L2C), UMR 5221 CNRS-Universit\'e de Montpellier, Montpellier,
  F-France.}

\begin{abstract}
Numerical data on scaling of the normalized Binder cumulant and the
normalized correlation length are shown for the Thermodynamic limit
regime, first for canonical Ising ferromagnet models and then for a
range of Ising spin glass models. A fundamental Renormalization Group
Theory rule linking the critical exponents for the two observables is
well obeyed in the Ising models, but not for the Ising spin glasses in
dimensions three and four. We conclude that there is a violation of
a standard Josephson hyperscaling rule in Ising spin glasses.
\end{abstract}


\maketitle

\section{Introduction}
The consequences of the Renormalization Group Theory (RGT) approach
have been studied in exquisite detail in numerous regular physical
models, typified by the canonical near neighbor interaction
ferromagnetic Ising model. It has been tacitly assumed that
Edwards-Anderson Ising Spin Glasses (ISGs), where the interactions are
random, follow the same basic scaling and Universality rules as the
regular models.

The Binder cumulant~\cite{binder:81} is an important observable which
has been almost exclusively exploited numerically for its scaling
properties as a dimensionless observable very close to criticality in
the finite-size scaling (FSS) regime $L \ll \xi(\beta)$, where $L$ is
the sample size and $\xi(\beta)$ the second-moment correlation length
at inverse temperature $\beta$. Here we will consider its scaling
properties in the Thermodynamic limit (ThL) regime $L \gg \xi(\beta)$
where the properties of a finite-size sample if normalized correctly
are independent of $L$ and so are the same as those of the
infinite-size model. (A standard rule of thumb for the onset of the
ThL regime is $L > 7 \xi(\beta,L)$).

In Ising ferromagnets, the second field derivative of the
susceptibility $\chi_{4}$ in a hypercubic lattice is directly related
to the Binder cumulant, see Eq.~{10.2} of Ref.~\cite{privman:91},
through
\begin{equation}
  2g(\beta,L) = \frac{-\chi_{4}}{L^d\chi^{2}} = \frac{3\langle
    m^2\rangle^2 - \langle m^4\rangle}{\langle m^2\rangle^2}
\end{equation}

The susceptibility $\chi$ scales with the critical exponent $\gamma$,
and the critical exponent for the second field derivative of the
susceptibility $\chi_{4}$ (also called the non-linear susceptibility),
is~\cite{butera:02}
\begin{equation}
  \gamma_{4}=\gamma +2\Delta_{\mathrm{gap}}  = d\nu + 2\gamma
  \label{gam4}
\end{equation}
because of very basic RGT scaling and hyperscaling~\cite{josephson:66}
relationships between exponents. $\nu$ and $\gamma$ are the standard
critical exponents for the correlation length and the susceptibility;
a textbook definition of hyperscaling is : "Identities obtained from
the generalised homogeneity assumption involve the space dimension
$d$, and are known as hyperscaling relations."~\cite{simons:97}.
Explicitly quoting Ref.~\cite{pelissetto:02} : "Below the upper
critical dimension, the following hyperscaling relations are supposed
to be valid: $2-\alpha = d\nu$, and $2\Delta_{\mathrm{gap}} = d\nu
+\gamma$ where $\Delta_{\mathrm{gap}}$ is the gap exponent, which
controls the radius of the disk in the complex-temperature plane
without zeroes, i.e. the gap, of the partition function (Yang-Lee
theorem)".

Thus in the ThL regime the normalized Binder cumulant $L^d g(\beta,L)$
(or alternatively $\chi_{4}(\beta)/(2\chi(\beta)^2)$) scales with the
critical exponent $(d\nu + 2\gamma) - 2\gamma = d\nu$, together with
correction terms as for any such observable.

For high temperatures, it can be noted that in any $S=\tfrac{1}{2}$
Ising system the infinite-temperature (i.e. independent spins) limit
for the Binder cumulant is $g(0,N) \equiv 1/N$, where $N$ is the
number of spins; $N=L^d$ for a hypercubic lattice so at infinite
temperature $L^{d}g(\beta,L) \equiv 1$.

Hyperscaling is well established in standard models, such as the Ising
models in dimensions less than the upper critical dimension, but many
years ago hyperscaling relations were predicted to break down in
quenched random systems~\cite{schwartz:91}. Two hyperscaling relations
have been quoted above; the first is well known and concerns the
specific heat exponent $\alpha$. The breakdown of this hyperscaling
relation in the Random Field Ising model has been extensively studied
\cite{gofman:93,vink:10}. In the Ising spin glasses which we discuss
below $\alpha$ is strongly negative and so is very hard to measure
directly; we will be concerned only with the second hyperscaling
relation which is less well known. We are aware of no tests of this
hyperscaling relation in a system with quenched randomness such as a
spin glass.

First we outline the general scaling approach, which follows
Ref.~\cite{campbell:06}. The Wegner scaling expression
\cite{wegner:72} for an observable $Q(\tau)$ in the ThL regime is
\begin{equation}
  Q(\tau) = C_{Q}\tau^{-\lambda}\left(1 + a_{Q}\tau^{\theta} + \cdots\right)
\end{equation}
where $\tau$ is a temperature scaling variable which tends to zero at
criticality $\beta=\beta_{c}$; $Q(\tau)$ diverges with the critical
exponent $\lambda$. In order to cover the entire paramagnetic
temperature range, the conventional RGT scaling variable $t =
(T-T_c)/T_c$ cannot be used as the temperature scaling variable
because it diverges at infinite temperature. Following Wegner and High
Temperature Scaling Expansion (HTSE) studies~\cite{butera:02,
  butera:03} it is appropriate to use for Ising models the alternative
temperature variable $\tau = 1-\beta/\beta_{c}$, which tends to $1$ at
infinite temperature. For Ising spin glasses (ISGs) having an
interaction distribution with symmetry between positive and negative
interactions, an appropriate scaling variable is $\tau =
1-(\beta/\beta_{c})^2$~\cite{daboul:04,campbell:06}. It is convenient
to normalise all observables $Q(\tau)$ in such a way that the infinite
temperature limit is $1$, not $0$ or $\infty$, as otherwise a
diverging set of correction terms would be required. The observables
$\chi(\tau,L)$ and $L^{d}g(\tau,L)$ automatically obey this
condition. The second moment correlation length $\xi(\tau,L)$ fulfills
the condition when normalized to $\xi(\tau,L)/\beta^{1/2}$ in Ising
models and to $\xi(\tau,L)/\beta$ in ISGs~\cite{campbell:06}. As will
be seen below, with this approach data can be fitted to high precision
over the entire paramagnetic temperature range from criticality to
infinity with a minimal set of Wegner correction terms (in principle
there is an infinite set of correction terms but with these
normalisations the higher order terms become extremely weak). Assuming
$\beta_{c}$ is known or can be estimated precisely, the true critical
exponents for the various observables can be estimated quite
accurately from ThL data without the need to approach $\beta_{c}$
closely.

\section{Ising models}

We will first study two canonical Ising models: the square lattice
model in dimension $2$ and the simple cubic model in dimension $3$, in
order to validate our method for testing hyperscaling.  We will choose
as the thermal scaling variable in the Ising models $\tau =
1-\beta/\beta_c$. $\beta_{c}$ is known exactly in dimension $2$ and to
very high precision in dimension $3$~\cite{haggkvist:07}. For the
simulation data the standard finite $L$ definition for the second
moment correlation length $\xi(\beta,L)$ through the Fourier
transformation of the correlation function has been used, see
Ref.~\cite{hasenbusch:08}, Eq. $14$. We will follow the "extended
scaling" convention introduced above and take the normalized
correlation length, $\xi(\beta,L)/\beta^{1/2}$.  The ThL normalized
correlation length behaves as
\begin{equation}
  \frac{\xi(\beta,L)}{\beta^{1/2}} = C_{\xi}\tau^{-\nu}\left(1 +
  a_{\xi}\tau^{\theta} + \cdots\right)
\end{equation}
It diverges with the standard critical exponent $\nu$ at criticality,
and tends to $1$ exactly at infinite temperature $\tau = 1$ for any
dimension $d$.  The term in $\tau^{\theta}$ is the leading Wegner
correction~\cite{wegner:72}. With the approach we follow, as ThL
correlation length corrections are relativly small for these Ising
models in either dimension $2$ or dimension $3$
\cite{campbell:08,campbell:11} the effective correlation length
exponent $\nu(\tau) =
\partial\ln[\xi(\beta,L)/\beta^{1/2}]/\partial\ln\tau$ is only weakly
temperature dependent from criticality to close to infinite
temperature. It is not necessary to go very close to criticality in
order to observe an effective exponent $\nu(\tau)$ very similar in
value to the true critical value. When simulation data for different
sizes $L$ are displayed together the ThL regime can be identified by
inspection as the regime where the observable becomes $L$
independent. A very detailed analysis of susceptibility, correlation
length and specific heat data for the $3$D Ising model following the
present approach is given in Ref.~\cite{campbell:11}.

Now consider the Binder cumulant in the Ising models. From the
relations above, the normalized ThL Binder cumulant $L^d g(\tau,L)$
has an infinite-temperature limit which is strictly $1$, and a ThL
regime behavior with a critical exponent $d\nu$ if hyperscaling holds,
together with Wegner corrections as for other observables, so:
\begin{equation}
  L^d g(\tau,L)\tau^{d\nu} = C_{g} \left(1+  a_{g}\tau^\theta + \cdots\right)
  \label{gscaled}
\end{equation}
Once again if the correction terms are relatively weak an
approximately assymptotic behavior, with an effective exponent very
similar to $d\nu$, will set in well before true criticality.
 
There is a hyperuniversal combination of critical amplitudes, the
\lq\lq dimensionless renormalized coupling constant\rq\rq{}
\cite{butera:02}, which with the present conventions then can be
written as $G_{r} = \beta_{c}^{d/2}C_{g}/C_{\xi}^{d}$.

So, in an Ising model if we plot the logarithm of the normalized
Binder cumulant $y(\tau)= \ln[L^{d}g(\tau,L)]$ and the logarithm of
the normalized correlation length $y(\tau) =
\ln[\xi(\tau,L)/\beta^{1/2}]$ against $x=\ln\tau$, the ThL negative
slopes $\partial y(\tau)/\partial\ln\tau$ will tend to $d\nu$ and
$\nu$ respectively as criticality is approached; for all $L$ the data
will tend to $y(\tau)=0$ at the infinite temperature $\tau = 1$ limit.

\begin{figure}
  \includegraphics[width=3.0in]{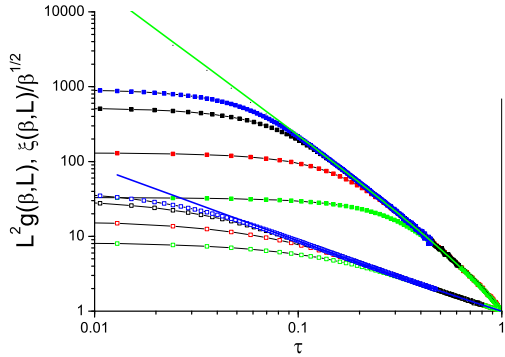}
  \caption{(Color on line) The $2$D ferromagnetic Ising model. Upper
    data sets $\ln[L^{2}g(\beta,L)]$ against $\tau$, lower data sets
    $\ln[\xi(\beta,L)/\beta^{1/2}]$ against $\ln\tau$ where
    $g(\beta,L)$ is the Binder cumulant and $\xi(\beta,L)$ is the
    second-moment correlation length.  Sizes $L = 32$, $24$, $12$,
    $6$, top to bottom. Smooth (upper) green curve and (lower) blue
    curve: fits (see text).} \protect\label{fig1}
\end{figure}

\begin{figure}
  \includegraphics[width=3.0in]{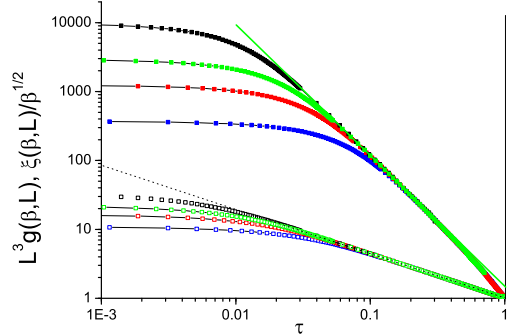}
  \caption{(Color on line) The $3$D ferromagnetic Ising model. Upper
    data sets $\ln[L^{3}g(\beta,L)]$ against $\ln\tau$, lower data
    sets $\ln[\xi(\beta,L)/\beta^{1/2}]$ against $\tau$ where
    $g(\beta,L)$ is the Binder cumulant and $\xi(\beta,L)$ is the
    second-moment correlation length.  Sizes $L = 24$, $16$, $12$,
    $8$, top to bottom. Smooth (upper) green curve and (lower) dashed
    curve: fits (see text)}\protect\label{fig2}
\end{figure}

For the Ising model ferromagnets on the $2$D square lattice and the
$3$D simple cubic lattice, data are displayed in this way in
Figs.~\ref{fig1} and \ref{fig2} respectively.  The ThL envelope where
data become independent of $L$ for large enough $L$ can be identified
by inspection.

Data for these plots are taken from numerical simulations for the $2$D
square lattice Ising model, for sample sizes $L= 6$, $12$, $24$, and
$32$, and in the $3$D data from numerical simulations simple cubic
Ising ferromagnet~\cite{haggkvist:07} for sample sizes $L= 8$, $12$,
$16$, and $24$. The ThL simulation curves agree in full detail with
data obtained by explicitly summing tabulated HTSE series
\cite{butera:02,butera:02a,butera:03}.

For these two Ising models all the relevant ThL parameters including
the critical exponents $\nu \equiv 1$ and $\nu = 0.6300$ respectively
and the critical amplitudes, are already known
\cite{butera:02,butera:03,simmons:15}.  For the $2$D Ising model with
the HTSE parameters, assuming hyperscaling, and keeping only the
leading correction terms with analytic correction term exponents
$\theta=1$, the calculated curves with no free parameters are
\begin{align}
  \frac{\xi(\tau)}{\beta^{1/2}} &= 0.854\tau{-1}\left(1 +0.171\tau\right)\\
  \frac{\chi_{4}(\tau)}{2\chi(\tau)^2} &= 2.365\tau{-2}\left(1 +0.577\tau\right)
\end{align}
which are in excellent agreement with the ThL simulation data
envelope, (where $L^{2}g(\tau,L)$ replaces
$\chi_{4}(\tau)/(2\chi(\tau)^2)$).

For the $3$D Ising model we again take the known critical temperature
and critical amplitudes.  While the effective leading correction term
exponent for the correlation length takes the standard value for this
model, $\theta=0.52$~\cite{campbell:08}, the optimal fit corresponds to
an effective $\theta_{\mathrm{eff}}$ for the HTSE $\chi_{4}$ data, and
hence for $L^{3}g(\tau,L)$, of $\theta_{\mathrm{eff}} \approx
1.4$. This can be ascribed for instance to the presence of a strong
higher order correction term with exponent $1+\theta$
\cite{butera}. The simulation data and the essentially exact HTSE data
(see Ref.~\cite{butera:02}, Fig.~19) can be fitted with the same large
effective exponent as the simulation data. The HTSE curves calculated
with the parameters
\begin{align}
  \frac{\xi(\tau)}{\beta^{1/2}} &= 1.074\tau{-0.63}\left(1-0.069\tau^{0.52}\right)\\
  \frac{\chi_{4}(\tau)}{2\chi(\tau)^2} &= 1.523\tau{-1.89}\left(1 -0.343\tau^{1.5}\right)
\end{align}
are in excellent agreement with the simulation ThL envelope data (with
$L^{3}g(\tau,L)$ replacing $\chi_{4}(\tau)/(2\chi(\tau)^2)$).

The $3$D ThL critical $L \equiv \infty, \tau \to 0$ limit for the
dimensionless renormalisation constant $G_{r} =
\beta_{c}^{d/2}C_{g}/C_{\xi}^{d}$ is $1.226$, which is quite different
from the finite size scaling $\tau \equiv 0, L \to \infty$ limit for
the same ratio which is $0.2803$.

In both models, over the entire temperature range covered by the
simulation data (excepting the region close to infinite temperature),
the almost assymptotic ratio of the $L^{d}g(\tau,L)$ and
$\xi(\tau,L)/\beta^{1/2}$ log-log slopes is very close to $d$,
verifiying that the hyperscaling rule is respected and incidentally
validating once again the utility of the extended scaling
normalization of the correlation length for Ising ferromagnet analyses
over wide temperature ranges~\cite{campbell:06a,campbell:11}. (A
"cross-over" behavior of high temperature data in Ising ferromagnets
\cite{luijten:97} is an artefact~\cite{lundow:11}).

\section {Ising spin glasses}

Now we turn to ISGs. The standard ISG Hamiltonian is $\mathcal{H}= -
\sum_{ij}J_{ij}S_{i}S_{j}$ with the near neighbor symmetric
distributions normalized to $\langle J_{ij}^2\rangle=1$. The
normalized inverse temperature is $\beta = (\langle
J_{ij}^2\rangle/T^2)^{1/2}$. The Ising spins live on simple
hyper-cubic lattices with periodic boundary conditions. The spin
overlap parameter is defined as usual by
\begin{equation}
  q = \frac{1}{L^{d}}\sum_{i} S^{A}_{i}S^{B}_{i}
\end{equation}
where $A$ and $B$ indicate two copies of the same system.  Klein {\it
  et al}~\cite{klein:91} quote exactly the same hyperscaling relation
Eq.~\eqref{gam4} for $\chi_{4}$ in the ISGs as in the Ising
ferromagnets (with the spin overlap moments $\langle q^2\rangle$ and
$\langle q^4\rangle$ replacing the magnetization moments $\langle
m^2\rangle$ and $\langle m^4\rangle$), so the RGT prediction for the
ISG Binder cumulant critical exponent is again $\gamma_{4}- 2\gamma =
d\nu$. Because the interaction parameter in the ISGs is $\langle
J_{ij}^2\rangle$ the appropriate ISG scaling variable is $\tau =
1-(\beta/\beta_{c})^2$~\cite{daboul:04,campbell:06} and the
appropriate normalized correlation length is $\xi(\tau,L)/\beta$
\cite{campbell:06}.

The simulation data in the ISGs are the same as those in
Refs.~\cite{lundow:15,lundow:15a} where the simulation techniques have
already been described in detail. Means were taken on at least $8192$
samples for each $L$ with of the order of $40$ different
temperatures. The maximum sizes studied were $L = 24$ for the bimodal
model and $L=16$ for the Gaussian model in dimension 3, $L = 14$ for
the bimodal model, $L = 12$ for the Gaussian and Laplacian models in
dimension $4$ and $L=10$ in dimension $5$.  Particular attention was
paid to achieving full equilibration; here we are only concerned with
data in the ThL above the ordering temperature where equilibration is
reached much faster than at or below the ordering temperature, so one
can have full confidence that the samples were in equilibrium for the
temperature range of interest here. For the $3$D bimodal model a
comparison with tabulated data generously provided as a supplement to
Ref.~\cite{hasenbusch:08} confirms equilibration.
  
\begin{figure}
  \includegraphics[width=3.0in]{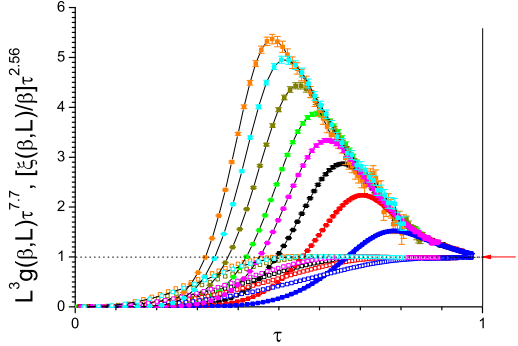}
  \caption{(Color on line) The $3$D bimodal ISG model. Upper data sets
    $L^{3}g(\beta,L)\tau^{7.7}$ against $\tau$, lower data sets
    $[\xi(\beta,L)/\beta]\tau^{2.56}$ against $\tau$ where
    $g(\beta,L)$ is the Binder cumulant and $\xi(\beta,L)$ is the
    second-moment correlation length.  Sizes $L = 24$, $20$, $16$,
    $12$, $10$, $8$, $6$, $4$ (upper set) and $L=24$, $20$, $12$, $6$
    (lower set), top to bottom.}\protect\label{fig3}
\end{figure}

\begin{figure}
  \includegraphics[width=3.0in]{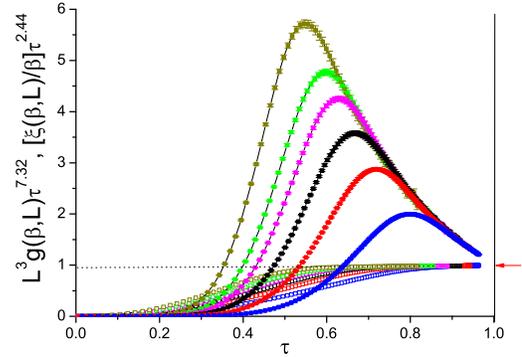}
  \caption{(Color on line) The $3$D Gaussian ISG model. Upper data
    sets $L^{3}g(\beta,L)\tau^{7.2}$ against $\tau$, lower data sets
    $[\xi(\beta,L)/\beta]\tau^{2.44}$ against $\tau$ where
    $g(\beta,L)$ is the Binder cumulant and $\xi(\beta,L)$ is the
    second-moment correlation length.  Sizes $L = 16$, $12$, $10$,
    $8$, $6$, $4$ (both sets), top to bottom.}\protect\label{fig4}
\end{figure}

We first display the data for the $3$D bimodal and Gaussian ISG models
in a form which provides a preliminary qualitative test of the
hyperscaling rule, see Figs.~\ref{fig3} and \ref{fig4}. We take the
most recent estimations for the critical parameters from the
literature : $\beta_{c} =0.909$ and $\nu = 2.56$ for the bimodal model
\cite{baity:13}, $\beta_{c} = 1.05$ and $\nu = 2.44$ for the Gaussian
model~\cite{katzgraber:06}. For each model we then plot together on
the same figure the products $[\xi(\tau,L)/\beta]\tau^{\nu}$ and
$L^{d}g(\tau,L)\tau^{d\nu}$ against $\tau$. If hyperscaling is
respected, we would expect that in each case both sets of ThL envelope
curves should lie close to $1$ for the whole range of $\tau$, to
within weak correction terms.  For both models the
$[\xi(\tau,L)/\beta]\tau^{\nu}$ ThL envelope curves indeed take this
form, but the $L^{d}g(\tau,L)\tau^{d\nu}$ ThL envelope curves (on the
right hand side of the observed peaks in the figures) behave in a
totally different manner suggestive of divergence with increasing
$L$. This is a dramatic qualitative demonstration that at least for
these two ISG models the hyperscaling rule is not obeyed.

\begin{figure}
  \includegraphics[width=3.0in]{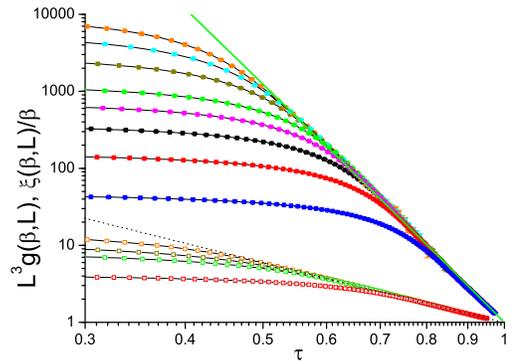}
  \caption{(Color on line) The $3$D bimodal ISG model. Upper data sets
    $\ln[L^{3}g(\beta,L)]$ against $\ln\tau$, lower data sets
    $\ln[\xi(\beta,L)/\beta]$ against $\ln\tau$ where $g(\beta,L)$ is
    the Binder cumulant and $\xi(\beta,L)$ is the second-moment
    correlation length.  Sizes $L = 24$, $20$, $16$, $12$, $10$, $8$,
    $6$, $4$ (upper set) and $L=24$, $20$, $12$, $6$ (lower set), top
    to bottom. Smooth (upper) green curve and (lower) dashed curve:
    fits (see text)}\protect\label{fig5}
\end{figure}

\begin{figure}
  \includegraphics[width=3.0in]{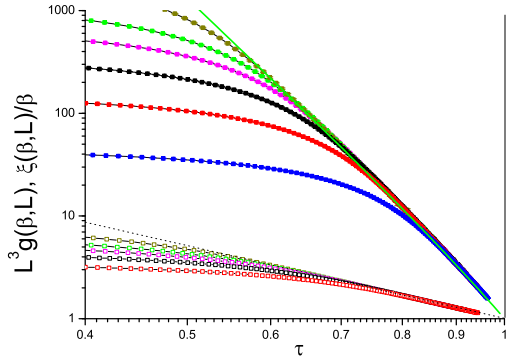}
  \caption{(Color on line) The $3$D Gaussian ISG model. Upper data
    sets $\ln[L^{3}g(\beta,L)]$ against $\ln\tau$, lower data sets
    $\ln[\xi(\beta,L)/\beta]$ against $\ln\tau$ where $g(\beta,L)$ is
    the Binder cumulant and $\xi(\beta,L)$ is the second-moment
    correlation length.  Sizes $L = 16$, $12$, $10$, $8$, $6$, $4$
    (upper set) and $L=16$, $12$, $10$, $8$, $6$ (lower set), top to
    bottom. Smooth (upper) green curve and (lower) dashed curve: fits
    (see text)}\protect\label{fig6}
\end{figure}

To obtain quantitative information, we will follow up by presenting
these ISG data in just the same way as in the figures for the Ising
models. Figs.~\ref{fig5} and \ref{fig6} show log-log plots of
$L^{3}g(\tau,L)$ against $\tau$ (upper data sets) and
$\xi(\tau,L)/\beta$ aginst $\tau$ (lower data sets) for the $3$D
bimodal ISG and the $3$D Gaussian ISG. We have assumed again for the
bimodal model $\beta_{c}=0.909$~\cite{baity:13} and for the Gaussian
model $\beta{c} = 1.05$~\cite{katzgraber:06}. The finite size scaling
correction exponent for the bimodal model has been estimated to be
$\omega \approx 1.1$~\cite{hasenbusch:08,baity:13} which corresponds
to a Wegner correction exponent $\theta = \nu\omega \approx
2.5$. There is no equivalent estimate available for the Gaussian model
so we will assume the same effective $\theta$.

Fits to the $\xi(\tau,L)/\beta$ ThL envelopes give effective exponents
$\nu = 2.5(1)$ for the bimodal ISG and $\nu = 2.4(1)$ for the Gaussian
ISG with negligible correction terms. These estimates, from
intermediate and high temperature data, are very similar to the
critical values quoted above estimated by finite size scaling, showing
that corrections to scaling are playing only a very minor role over
the whole paramagnetic temperature range.  The fits for the
$L^{3}g(\tau,L)$ ThL envelopes are
$1.2\tau^{-10.05}(1-0.17\tau^{2.5})$ for the bimdal model and
$2.0\tau^{-9.5} (1-0.50\tau^{2.5})$ for the Gaussian model,
i.e. critical exponent estimates which are $4.00(5)\nu$ and
$3.96(5)\nu$ respectively, so well above the hyperscaling value of
$3\nu$.

\begin{figure}
  \includegraphics[width=3.0in]{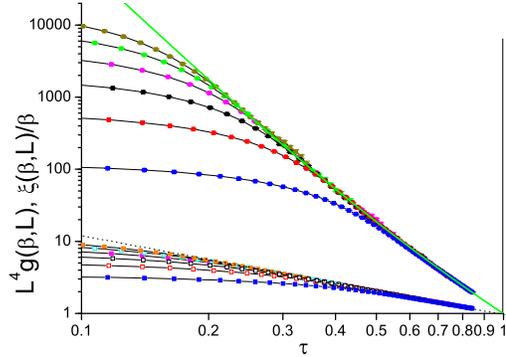}
  \caption{(Color on line) The $4$D bimodal ISG model. Upper data sets
    $\ln[L^{4}g(\beta,L)]$ against $\ln\tau$, lower data sets
    $\ln[\xi(\beta,L)/\beta]$ against $\ln\tau$ where $g(\beta,L)$ is
    the Binder cumulant and $\xi(\beta,L)$ is the second-moment
    correlation length.  Sizes $L = 14$, $12$, $10$, $8$, $6$, $4$
    (both sets), top to bottom. Smooth (upper) green curve and (lower)
    dashed curve: fits (see text)}\protect\label{fig7}
\end{figure}

\begin{figure}
  \includegraphics[width=3.0in]{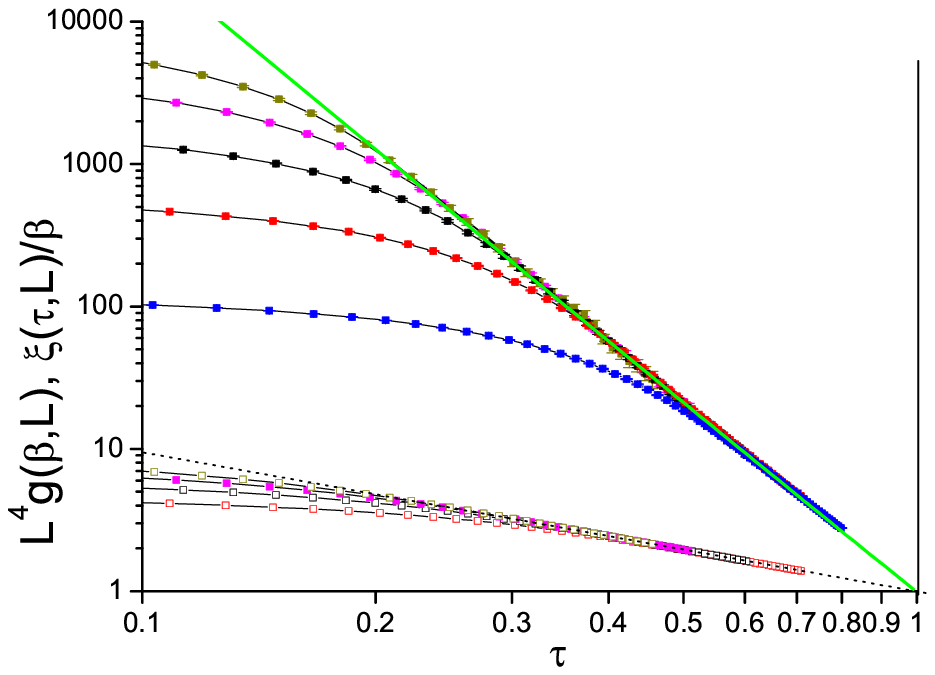}
  \caption{(Color on line) The $4$D Gaussian ISG model. Upper data
    sets $\ln[L^{4}g(\beta,L)]$ against $\ln\tau$, lower data sets
    $\ln[\xi(\beta,L)/\beta]$ against $\ln\tau$ where $g(\beta,L)$ is
    the Binder cumulant and $\xi(\beta,L)$ is the second-moment
    correlation length.  Sizes $L = 12$, $10$, $8$, $6$, $4$ (upper
    set) and $L=12$, $10$, $8$, $6$ (lower set), top to bottom. Smooth
    (upper) green curve and (lower) dashed curve: fits (see
    text)}\protect\label{fig8}
\end{figure}

\begin{figure}
  \includegraphics[width=3.0in]{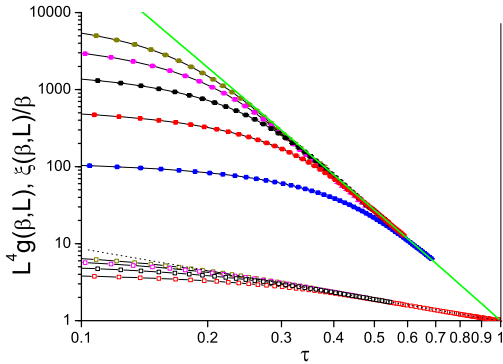}
  \caption{(Color on line) The $4$D Laplacian ISG model. Upper data
    sets $\ln[L^{4}g(\beta,L)]$ against $\ln\tau$, lower data sets
    $\ln[\xi(\beta,L)/\beta]$ against $\ln\tau$ where $g(\beta,L)$ is
    the Binder cumulant and $\xi(\beta,L)$ is the second-moment
    correlation length.  Sizes $L = 12$, $10$, $8$, $6$, $4$ (upper
    set) and $L=12$, $10$, $8$, $6$ (lower set), top to bottom. Smooth
    (upper) green curve and (lower) dashed curve: fits (see
    text)}\protect\label{fig9}
\end{figure}

Continuing on to dimension $4$, Figures~\ref{fig7}, \ref{fig8} and
\ref{fig9} show data presented in the same manner, for the bimodal,
Gaussian and Laplacian $4$D ISG models. The critical inverse
temperatures are $\beta_{c}=0.505$ for the bimodal model
\cite{lundow:15}, $\beta_{c}=0.555$ for the Gaussian model
\cite{jorg:08,lundow:15} and $\beta_{c}=0.419$ for the Laplacian model
\cite{lundow:16}.  Fits to the ThL envelopes give effective exponents
$\nu = 1.10$ for the bimodal, $\nu = 1.02$ for the Gaussian and $\nu =
0.95$ for the Laplacian model. These estimates from intermediate and
high temperature data are very similar to the critical values
estimated by finite size scaling, $\nu= 1.12$, $\nu =1.02$
\cite{lundow:15} and $\nu= 0.99$~\cite{lundow:16} respectively showing
that corrections to scaling at temperatures well above $T_{c}$ are
again playing only a very minor role.  The fit to the ThL envelope of
the bimodal $L^{4}g(\tau,L)$ data set is $L^{4}g(\tau,L)=
0.22\tau^{-5.5}(1+3.5\tau^{2.0})$ so with a very strong correction to
scaling. The fit to the Gaussian ThL envelope is $L^{4}g(\tau,L) =
0.9\tau^{-4.5}(1+0.1\tau^{1.5})$ and the fit to the Laplacian ThL
envelope is $L^{4}g(\tau,L) = 1.2\tau^{-4.6}(1-0.17\tau^{1.5})$ so
both with rather weak corrections to scaling. The $L^{4}g(\tau,L)$
critical exponent estimates are $5.0(1)\nu $, $4.4(1)\nu$ and
$4.65(10)\nu$ respectively so well above the hyperscaling prediction,
for which the values should be equal to $4\nu$.

\begin{figure}
  \includegraphics[width=3.0in]{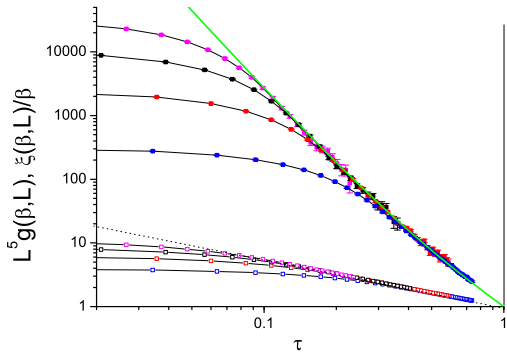}
  \caption{(Color on line) The $5$D bimodal ISG model. Upper data sets
    $\ln[L^{5}g(\beta,L)]$ against $\ln\tau$, lower data sets
    $\ln[\xi(\beta,L)/\beta]$ against $\ln\tau$ where $g(\beta,L)$ is
    the Binder cumulant and $\xi(\beta,L)$ is the second-moment
    correlation length.  Sizes $L = 10$, $8$, $6$, $4$, top to bottom
    in each set. Smooth (upper) green curve and (lower) dashed curve:
    fits (see text)}\protect\label{fig10}
\end{figure}

\begin{figure}
  \includegraphics[width=3.0in]{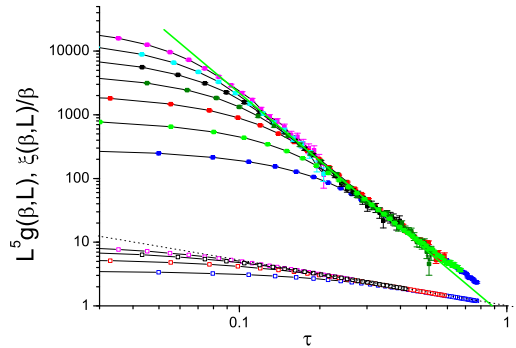}
  \caption{(Color on line) The $5$D Gaussian ISG model. Upper data
    sets $\ln[L^{5}g(\beta,L)]$ against $\ln\tau$, lower data sets
    $\ln[\xi(\beta,L)/\beta]$ against $\ln\tau$ where $g(\beta,L)$ is
    the Binder cumulant and $\xi(\beta,L)$ is the second-moment
    correlation length.  Sizes $L = 10$, $9$, $8$, $7$, $6$, $5$, $4$
    (upper set) and $L=10$, $8$, $6$, $4$ (lower set), top to
    bottom. Smooth (upper) green curve and (lower) dashed curve: fits
    (see text)}\protect\label{fig11}
\end{figure}

Finally in dimension $5$ the bimodal and Gaussian ISG inverse critical
temperatures are $\beta_c = 0.3885$ and $\beta_c = 0.419$
\cite{lundow:16}. The same procedure is followed as for the other
dimensions, Figures~\ref{fig10} and \ref{fig11}.

Fits to the $\xi(\tau,L)/\beta$ ThL envelopes give effective exponents
$\nu = 0.75(1)$ for the $5$D bimodal ISG and $\nu = 0.76(1)$ for the
$5$D Gaussian ISG with negligible correction terms. These estimates,
from intermediate and high temperature data, are similar to the
critical values estimated by finite size scaling, $\nu = 0.77(2)$,
$\nu=0.72(1)$ respectively~\cite{lundow:16}, showing that corrections
to scaling are playing only a minor role over the whole paramagnetic
temperature range for this observable.  The fits for the
$L^{5}g(\tau,L)$ ThL envelopes are $0.20\tau^{-4.1}(1+4.0\tau^{1.7})$
for the bimodal model and $0.50\tau^{-3.6} (1+1.0\tau^{1.0})$ for the
Gaussian model, i.e. critical exponent estimates which are $5.4(2)\nu$
and $4.9(2)\nu$ respectively. The corrections to scaling are very
strong and the correction exponents $\theta_{\mathrm{eff}}$ are not
accurately determined so the uncertainties are stronger than for $3$D
and $4$D ; for $5$D the data indicate an exponent ratio which is
compatible with or weakly above the hyperscaling value of $5\nu$.

\section{Conclusion}

From RGT, the critical exponent for the normalised Binder cumulant in
dimension $d$, $\chi_{4}(\tau,L)/(2\chi(\tau,L)^{2}) = L^{d}g(\tau,L)$
should be strictly equal to $d\nu$ in all models below the upper
critical dimension. However, the scaling relation is based on the
assumption that the standard Josephson hyperscaling rule
Eq.~\ref{gam4} holds. The derivation leading to this rule assumes
translational invariance (see \cite{schwartz:91,muller:15}). It is
well established that in the Random Field Ising model (RFIM), which is
not translationally invariant, the hyperscaling rule $2-\alpha = \beta
+2\gamma = \nu d$ is replaced by $2-\alpha = \beta +2\gamma =
\nu(d-\theta)$ ; here $\theta$ is a hyperscaling violation exponent
(not to be confused with the correction exponent) and it is believed
that $\theta=\gamma/\nu$ (two exponent scaling) in the RFIM~\cite
{gofman:93,vink:10}. ISGs are not translationally invariant either. In
ISGs $\alpha$ is always negative and very large, and so is very hard
to estimate directly. The ISG Binder cumulant scaling behavior
reported here provides another test of hyperscaling, for a Binder
cumulant hyperscaling relation on seven different ISG models.

We conclude empirically from the simulation data that the basic RGT
scaling law with the second Josephson hyperscaling rule does not hold
in ISGs, at least in dimensions $3$ and $4$, with dimension $5$ being
uncertain. It may be relevant that for spin glasses the situation
concerning the gap exponent $\Delta_{\mathrm{gap}}$ referred to above
is more complicated than for ferromagnets, or even for diluted
ferromagnets, as the locations of Yang-Lee zeros are not restricted to
the imaginary-field axis~\cite{matsuda:10}.

It has been generally accepted that RGT implies that in ISGs critical
exponents and the values of dimensionless parameters at criticality
should be universal, whatever the form of the interaction
distribution. It may be relevant that numerically this RGT
universality rule also has been found not to hold, for ISGs in $4$D
\cite{lundow:15,lundow:15a}, $2$D~\cite{lundow:15b} and $5$D
\cite{lundow:16}.  However, in the RFIM universality has been shown to
hold~\cite{fytas:13}. The link between non-universality and
hyperscaling breakdown is not clear to us. However it appears that
none of the standard RGT rules should be taken for granted in ISGs.

It has been authoritatively stated that \lq\lq classical tools of RGT
analysis are not suitable for spin glasses\rq\rq{}
\cite{parisi:01,castellana:11,angelini:13}. The numerical results
taken together indeed appear to be a clear empirical indication that a
fundamentally novel theoretical approach is required for scaling at
spin glass transitions.

\begin{acknowledgments}
  We would like to thank Professor A. Aharony, Dr. P. Butera and
  Dr. C. M\"{u}ller for helpful comments.  The computations were
  performed on resources provided by the Swedish National
  Infrastructure for Computing (SNIC) at the High Performance
  Computing Center North (HPC2N) and Chalmers Centre for Computational
  Science and Engineering (C3SE).
\end{acknowledgments}

\end{document}